# Predicting Likely-Vulnerable Code Changes: Machine Learning-based Vulnerability Protections for Android Open Source Project

**Keun Soo YIM, Senior Member, IEEE**
Google, Mountain View, CA 94043, USA

Corresponding author: K. S. Yim (e-mail: yim@google.com).

**ABSTRACT** This paper presents a framework that selectively triggers security reviews for incoming source code changes. Functioning as a review bot within a code review service, the framework can automatically request additional security reviews at pre-submit time before the code changes are submitted to a source code repository. Because performing such secure code reviews add cost, the framework employs a classifier trained to identify code changes with a high likelihood of vulnerabilities. The online classifier leverages various types of input features to analyze the review patterns, track the software engineering process, and mine specific text patterns within given code changes. The classifier and its features are meticulously chosen and optimized using data from the submitted code changes and reported vulnerabilities in Android Open Source Project (AOSP). The evaluation results demonstrate that our Vulnerability Prevention (VP) framework identifies approximately 80% of the vulnerability-inducing code changes in the dataset with a precision ratio of around 98% and a false positive rate of around 1.7%. We discuss the implications of deploying the VP framework in multi-project settings and future directions for Android security research. This paper explores and validates our approach to code change-granularity vulnerability prediction, offering a preventive technique for software security by preemptively detecting vulnerable code changes before submission.

## I. INTRODUCTION

The free and open source software (FOSS) supply chains for the Internet-of-Things devices (e.g., smartphones and TVs) present an attractive, economic target for security attackers (e.g., supply-chain attacks [20][21][28]). It is for instance because they can submit seemingly innocuous code changes containing vulnerabilities without revealing their identities and motives. The submitted vulnerable code changes can then propagate quickly and quietly to the end-user devices. Targeting specific, widely used open source projects (e.g., OS kernels, libraries, browsers, or media players) can maximize the impact, as those projects typically underpin a vast array of consumer products. The fast software update cycles of those products can quickly take vulnerabilities in the latest patches of their upstream FOSS projects if rigorous security reviews and testing are not implemented before each software update or release. As a result, those vulnerable code changes can remain undetected and thus unfixed, reaching a large number of end-user devices.

From a holistic societal perspective, the overall security testing cost can be optimized by identifying such vulnerable code changes early at pre-submit time, before those changes are submitted to upstream, open source project repositories. Otherwise, the security testing burden is multiplied across all the downstream software projects that depend on any of the upstream projects. Those downstream projects cannot rely on the first downstream projects to find and fix the merged, upstream vulnerabilities because the timeframe for such fixes and their subsequent upstreaming is unpredictable (e.g., in part due to the internal policies [22]). Thus, it is desirable to prevent vulnerable code submissions in the upstream projects. A naïve approach of requiring comprehensive security reviews for every code change cause an unrealistic cost for many upstream open source project owners. It is especially true for FOSS projects receiving a high volume of code changes or requiring specialized security expertise for reviews (e.g., specific to the domains).

To this end, this paper presents a *Vulnerability Prevention* (VP) framework that automates vulnerability assessment of code changes using a *machine learning (ML) classifier*. The classifier model estimates the likelihood that a given code change contains or induces at least one security vulnerability. Code changes exceeding a threshold mean *likely-vulnerable*. The model is trained on the historical data generated by using a set of associated *analysis tools*. The model uses the common features used for software defect prediction as well as *four types of novel features* that capture: (1) the patch set complexity, (2) the code review patterns, (3) the software development lifecycle phase of each source code file, and (4) the nature of a code change, as determined by analyzing the edited source code lines. In total, this study comprehensively examines 6 types of classifiers using over 30 types of feature data to optimize the accuracy of the ML model.

To generate the training and test data, we leverage the security bugs discovered and fixed in the Android Open Source Project (AOSP)[1]. It specifically targets the AOSP media project[2] (i.e., for multimedia data processing) that was extensively fuzz-tested and thus revealed many security defects. A set of specialized tools is designed and developed as part of this study to: (1) identify vulnerability-fixing change(s) associated with each target security bug, and (2) backtrack vulnerability-inducing change(s) linked to each of the identified vulnerability-fixing changes. All the identified vulnerability-inducing changes are then manually analyzed and verified before being associated with the respective security bugs. The associated vulnerability-inducing changes are labeled as '1', while all the other code changes submitted to the target media project are labeled as '0' in the dataset.

*The N-fold evaluation* using the first year of data identifies random forest as the most effective classifier based on its accuracy. The classifier identifies ~60% of the vulnerability-inducing code changes with a precision of ~85%. It also identifies ~99% of the likely-normal code changes with a precision of ~97% when using all the features for the training and testing.

The VP framework is then used as an *online model* retrained monthly on data from the previous month. When it is applied to about six years of the vulnerability data[3], the framework demonstrates an approximately 80% recall and an approximately 98% precision for vulnerability-inducing changes, along with a 99.8% recall and a 98.5% precision for likely-normal changes. This accuracy result surpasses the results achieved in the N-fold validation in large part because the online deployment mode can better utilize the underlying temporal localities, casualties, and patterns within the feature data.

In summary, 7.4% of the reviewed and merged code changes are classified as vulnerability-inducing. On average,

the number of likely-normal changes requiring additional attention during their code reviews is around 7 per month. This manageable volume (less than 2 code changes per week) justifies the cost, considering the high recall (~80%) and precision (~98%) for identifying vulnerability-inducing changes. The main contributions of this study include:

- We explore and confirm the possibility of code change-granularity vulnerability prediction that can be used to prevent vulnerabilities by flagging likely-vulnerable code changes at pre-submit time.
- We present the Vulnerability Prevention (VP) framework that automates online assessment of software vulnerabilities using a machine learning classifier.
- We devise novel feature types to improve the classifier accuracy and reduces the feature data set by evaluating the precision and recall metrics.
- We present the specialized tools to label code changes in AOSP, facilitating robust training and testing data collection.
- We demonstrate a high precision (~98%) and recall (~80%) of the VP framework in identifying vulnerability-inducing changes, showing the potential as a practical tool to reduce security risks.
- We discuss the implications of deploying the VP framework in multi-project settings. Our analysis data suggests two focus areas for future Android security research: optimizing the Android vulnerability fixing latency and more efforts to prevent vulnerabilities.

The rest of this paper is organized as follows. Section II provides the background information. Section III analyzes the design requirements and presents the VP framework design. Section IV details the design of the ML model, including the classifier and features for classifying likely-vulnerable code changes. Section V describes the tools developed to collect vulnerability datasets for model training and testing. Section VI describes the data collection process using the tools, and characterizes the vulnerability issues, vulnerability-fixing changes, and vulnerability-inducing changes in an AOSP sub-project. Section VII presents the evaluation of the VP framework using an N-fold validation. Section VIIII extends the framework for real-time, online classification. Section IX discusses the implications and threats to validity. Section IX reviews the related works before concluding this paper in Section X.

## II. BACKGROUND

This section outlines the code review and submission process of an open source software project, using AOSP (Android

---

Open Source Project) as a case study. AOSP is chosen, considering its role as an upstream software project with the significant reach, powering more than 3 billion, active end-user products.

**Code Change.** A code change (simply, change) consists of a set of added, deleted, and/or edited source code lines for source code files in a target source code repository (e.g., git). A typical software engineer sends a code change to a code review service (e.g., Gerrit[4]) for mandatory code reviews prior to submission. A code change is attributed to an author who has an associated email address in AOSP. The change can also have one or more code reviewers. Both the author and reviewers have specific permissions within each project (e.g., project ownership status and review level).

During the code review process, a code change can undergo multiple revisions, resulting in one or more patch sets. Each patch set uploaded to the code review service represents an updated version of the code change. The final, approved patch set of the change can then be submitted and merged into the target source code repository.

**Code Review.** The code change author can revise and resend the change as a new patch set for further review or approval by designated code reviewer(s). The key reviewer permissions include: a score of +1 to indicate the change looks good to the reviewer, a score of +2 to approve the code change, a score of -1 to tell that the change does not look good (e.g., a minor issue), and a score of -2 to block the code change submission. Projects (e.g., git repositories or sub-directories in a git repository) can have custom permissions and review rules. For example, a custom review rule is to enable authors to mark their code changes ready for pre-submit testing because often authors upload non-final versions to the code review service (e.g., to inspect the diffs[5] and preliminary feedback).

## III. DESIGN

This section outlines the design of the VP (Vulnerability Prevention) framework. The design is based on an analysis of its essential design requirements.

### A. DEFINITIONS

Let us define the secure code review points and a taxonomy for classifying code changes in this study.

**Secure Code Review Points.** It shows the three types of events that can be used to automatically trigger our classifier: (1) a code change is initially sent for code review (or marked as ready for review or pre-submit testing); (2) a new patch set is sent; (3) and a code change is submitted. Its use can be refined by extra conditions (e.g., triggering only when a reviewer is specified). The classifier can also be manually triggered in several ways (e.g., by adding a tag to the change description, clicking a UI button or checkbox in the code review service, or executing a shell command for a specific code change available in the code review service).

**Classification of Code Changes.** In this study, we classify code changes into the following categories:

- *ViC (Vulnerability-inducing Change)* for a code change that originally induced a vulnerability.
- *VfC (Vulnerability-fixing Change)* for a code change that fixed the existing vulnerability.
- *LNC (Likely Normal Change)* for a code change unlikely to induce a vulnerability. Notably, it includes changes that have not be identified as a known ViC at the time of analysis.

Additionally, *VfLs (Vulnerability-fixing Lines)* are the specific subset of source code lines edited by a VfC where the edits are essential to resolving the vulnerability.

### B. DESIGN GOALS

Our approach is devised for use cases meeting the following conditions:

- The target project experiences frequent software vulnerabilities with high potential consequences (e.g., costly fixes, product reputational damage, and impact on users).
- The target project serves as an upstream source for downstream software projects used to build many integrated, end-user products or services (e.g., AOSP for Android smartphones and TVs).
- Downstream projects often lack rigorous security testing (e.g., system fuzzing with dynamic analyzers [19]) due to the associated cost, technical expertise, and tooling constraints.

By detecting and blocking vulnerable code changes in the upstream target project, security engineering costs are reduced for downstream projects. The reduction encourages continued use of the upstream project and attracts additional downstream adoption, incentivizing the upstream project owners to invest in vulnerability prevention practices.

Under the targeted conditions, a classifier that estimates the likelihood of a vulnerability in a given code change proves effective. When the estimated likelihood exceeds a threshold, the respective code change is flagged for further scrutiny via secure code review or rigorous security testing. Our approach facilitates the detection of vulnerable code changes (e.g., with failing security tests) and consequently prevents their integration into the repository.

Our approach also offers seamless integration into post-submit, secure code review processes. Code changes flagged by the classifier undergo additional offline review by security engineers or domain experts. It increases the likelihood of vulnerability detection within those changes. By applying the classifier post-submission but pre-release, targeted security reviews can focus on the highest risk code changes (based on the estimated likelihood). Our approach,

---

thus, optimizes the secure code review process, reducing the overall security costs, while maintaining a robust security posture.

Considering those use cases, the key design goals for the classifier are set as follows:

- **Reasonable recall (>75% for ViCs).** It ensures that a significant majority (>75%) of vulnerable code changes is detected, substantially reducing the overall security risk.
- **High precision (>90% for ViCs).** Since vulnerable code changes are rare, incorrectly flagging 1 out of 10 code changes is generally acceptable for the security reviewers.
- **Low false positive ratio (<2%).** Normal code changes should rarely be flagged in order to maintain a streamlined review process for developers.
- **Fast inference time (e.g., <1 minute).** It is to enable smooth integration into a code review service without causing developer-visible delays.
- **Low inference cost.** To operate within typical open source project budgets for security infrastructure and tools, the inference should be done without having to use powerful or specialized ML hardware devices.
- **Infrequent retraining.** Because the cost for model retraining is also important, monthly retraining on up to about a million samples is considered acceptable. It is to balance the accuracy maintenance (vs. daily retraining) with the affordability for most open source projects.

### C. FRAMEWORK

Our VP framework is applicable to three distinct use cases:

- **Pre-submit Security Review** is to utilize VP for assessing every code change sent for code review and identify likely-vulnerable code changes for additional secure code review by security domain experts.
- **Pre-submit Security Testing** is to employ VP for assessing every code change sent for code review and identify likely-vulnerable code changes for extra security testing (e.g., static analysis, dynamic analysis, or fuzzing) before submissions.
- **Post-submit Security Review** is to apply VP to all code changes submitted within a predefined period (e.g., daily or weekly) and isolate a set of likely-vulnerable code changes for an additional in-depth a secure code inspection by security domain experts. This use case differs from the existing post-submit time, security testing.

The pre-submit security review use case scenario has the highest complexity. Compared with post-submit use cases, pre-submit use cases have stricter requirements for inference time and online retraining (e.g., directly visible to code change authors vs. quality assurance team). Compared with the pre-submit security testing use case, the pre-submit security review use case has stricter accuracy requirements

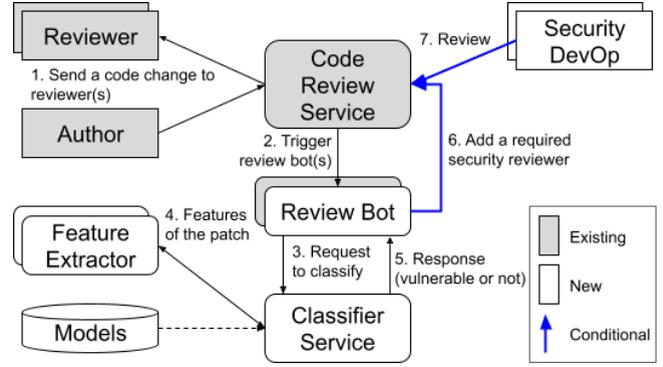

**FIGURE 1.** Overview of the Vulnerability Prevention (VP) Framework.

(e.g., false positives for security testing mean mostly extra testing costs). Thus, it is used as the primary target for the framework design. To address the selected use case, the VP (Vulnerability Prevention) framework leverages its following key subsystems as depicted in Figure 1:

**Code Review Service.** Authors initiate the code review process by uploading their code changes to a designated code review service. They then assign reviewers for their changes (see step 1 in Figure 1). The review service automatically triggers one or more review bots in response to the request of an author or reviewer, or when the uploaded code changes satisfy predefined conditions.

**Review Bot(s).** Triggered review bots access the specific edits made by a source code change, along with relevant metadata of the code change and the baseline source code. To conduct in-depth analysis, bots usually leverage backend services for compilation, analysis, and testing of the change against the baseline. The new VP review bot utilizes those capabilities and forwards the gathered data to the classifier service. The classifier service then determines if the given source code change is likely-vulnerable or not.

**Classifier Service.** When the classifier service is triggered, it utilizes the feature extractors and the data from the VP review bot to extract the pertinent features of the given code change. Subsequently, it performs inference using a model in order to estimate how likely the code change has vulnerabilities. The classifier model uses the extracted features as input and generates a binary output signal (i.e., '1' indicates likely-vulnerable and '0' indicates likely-normal). The output signal guides if additional security review (or security testing) is beneficial for the code change. The classifier service has an option to employ multiple models, combining their results (e.g., through logical operators or majority voting) for better accuracy.

**Notification Service.** When a code change is classified as a likely-vulnerable change, the notification service posts a comment on the code change in the code review service. The comment alerts the code change author and existing reviewers to the potential presence of vulnerabilities, urging extra scrutiny.

If the target project maintains a dedicated pool of security reviewers, the notification service can automatically assign a member of the pool as a required reviewer for the target code change. The selected reviewer can be a primary engineer on a rotation or chosen through heuristics (e.g., round robin) for balanced distribution of security review workload.

### D. EXTENSIONS
The VP framework supports extension for security testing and post-submit use cases:

**Selective Security Testing.** To extend the VP framework for selective, pre-submit security testing, an asynchronous test execution service is further employed. The execution service patches the given code change into the baseline code, builds artifacts (e.g., binaries), and executes relevant security tests against the build artifacts.

The execution service supports customization of security test configurations, including parameter tuning to target specific functions and adjust the maximum testing time. In implementation, a review bot is extended to generate tailored testing parameters. The extended bot leverages both the source code delta of the target code change and the vulnerability statistics of the target project. The resulting data-driven method allows the bot to use either the default parameter values or dynamically generate new ones, helping to optimize the balance between the security testing coverage and associated costs.

**Post-Submit Use Case.** To extend the VP framework for post-submit time use cases, a replay mechanism is needed to process submitted code changes and invoke the VP review bot with the relevant input data. In particular, it requires tracking the code change identifiers from the git commit hashes if the used version control system is git. It uses the classification results to select a subset of code changes for further comprehensive security review.

## IV. MODELING
This section presents the design of a machine learning (ML) model trained to classify likely-vulnerable code changes. It explores lightweight classifier options and describes both the new and established input feature data types.

### A. CLASSIFIERS
This study explores the following six common classifiers used for software fault prediction:

- **Decision Tree.** It is a kind of flowchart that captures the modeled decision making process where nodes represents decision points and branches represents outcomes.
- **Random Forrest.** It aggregates results from multiple decision trees to produce a single, enhanced result.

- **SVM (Support Vector Machine).** It is a supervised ML algorithm that finds an optimal hyperplane for the training data to separate test data points into distinct classes.
- **Logistic Regression.** It is a statistical method to model the log-odds of an event as a linear combination of one or more independent variables.
- **Naïve Bayes.** It models the distribution of inputs of a given class, assuming the input features are conditionally independent, given the target class.
- **Quinlan's C4.5 (v8)** [13]. It is an efficient algorithm to generate a decision tree.

The hypothesis is that those common classifiers would be effective in predicting likely-vulnerable code changes. To test the hypothesis, this study evaluates all the six classifiers. All the classifiers are selected for their lightweight nature (e.g., low training cost and short inference time), meeting the two key design requirements. This study seeks to identify which of those classifiers can also achieve the required level of classification accuracy.

### B. FEATURES
The classifier relies on a well-selected set of input feature data. The feature data types used in this study are as follows:

**Human Profile (HP).** The HP features capture the affiliations of the author and reviewer(s) of a code change:

- $HP_{author}$ represents the trustworthiness of the email domain of an author. Email domains are ranked on an integer scale starting with 1 for the most trustworthy domain type and increasing by 1 as the trustworthiness declines. In Android Open Source Project (AOSP), verified email domains are available, helping to capture the author organizations[6]. Here, the value of '1' is used when an author email domain is for the primary sponsor of AOSP (i.e., google.com); '2' is used when the domain is android.com; '3' is used for an Android partner company (e.g., samsung.com or qualcomm.com); '4' is for other relevant open source communities (e.g., kernel.org); and '5' is for all other domains (e.g., github.com or gmail.com).
- $HP_{reviewer}$ similarly represents the trustworthiness of the code reviewer organizations, using the same value scale as $HP_{author}$. It considers reviewers giving scores of '+2', '+1', or '-1' to a given code change. Code changes with a '-2' score from any reviewer is not submittable until the author addresses the respective review. Thus, the score of '-2' is excluded from the modeling. For each relevant reviewer, $HP_{reviewer}$ is calculated and then the largest value is taken (i.e., the most external reviewer organization). It is assumed that external reviewers exhibit different behavioral patterns compared to internal reviewers in identifying potential vulnerabilities (e.g., due to limited access to project-specific, internal information, while some external reviewers possess unique expertise in security or specific domains).

---

[6] AOSP uses the Gerrit code review service that employs plugin based validation mechanisms for new commits, new groups, account activations, review comments, pre-merge events, and on submit events.

While considering the roles of a code change author and reviewers (e.g., such as committer, active developer, tester, or release engineer) may seem useful, roles are not used in the modeling. It is because attackers can exploit trusted roles (e.g., as an attack vector). Furthermore, roles gleaned from the commit history do not reliably indicate the level of security expertise associated with each role.

**Change Complexity (CC).** The CC features represent the complexity of a given code change. It relies on a common observation that more complex change are more prone to software defects and thus to software vulnerabilities than the simpler changes. The following two CC features are used to gauge the likelihood of code change author mistakes:

- $CC_{add}$ counts the total number of lines added by a code change. It includes edits to all non-binary, text files in a code change such as source code, configuration files, and build files.
- $CC_{del}$ counts the total number of lines deleted by a code change in a similar manner.

Here, we note that a modified source code line is counted as both a deleted line and an added line.

Those two basic *change complexity* metrics directly measure the volume of a code change. They are simpler than the common *code complexity* metrics. Furthermore, unlike our code change complexity metrics, the existing code complexity metrics (e.g., Halstead complexity [2], McCabe complexity, or CK metrics [17]) are usually for a software module or equivalent but not for the delta made by a code change. Halstead complexity uses the number of operators and operands in a given software module. McCabe complexity basically gives more weights to the edges in a control flow graph of a given module than the nodes. It also has definitions for cyclomatic, essential, and design complexities. CK metrics are for object-oriented programs as they use the sum of the complexity of the methods of a class.

While the existing code complexity could be applied to measure the complexity of a change, it involves a complex process. For example, one may perform such a code complexity analysis twice before and after applying a given code change to the baseline code and then compute the delta of the calculated code complexity metric values. In practice, aggregating such the delta values of code complexity for various directories, file types, files, modules, classes, and functions is non-trivial. Thus, it alones warrants a dedicated study.

While complex code changes are inherently difficult to find vulnerabilities in, they often get extra attentions from the reviewers (e.g., more questions and more revisions). For example, complex, thoroughly-reviewed code changes might actually be safer than medium-complexity changes that received minimal reviews (e.g., rubberstamped). To model the degree of reviewer engagement, the Patch set Complexity (PC) and Review Pattern (RP) features are devised:

**Patch set Complexity (PC).** A code change has multiple patch sets if it undergoes multiple revisions (e.g., in response to a code review). The following PC features are specifically devised to capture the volume of those patch sets:

- $PC_{count}$ is the total number of patch sets uploaded before a given code change is finally merged to the repository.
- $PC_{revision}$ is the sum of the total number of source code lines added or deleted by each of the revised patch sets of a code change, excluding the first patch set. It thus captures the volume of revisions made since the first patch set. Here, the lines added or deleted by each patch set are determined by calculating the deltas (or differences) between consecutive patch sets.
- $PC_{relative\_revision}$ is a ratio of $PC_{revision}$ and the number of added or deleted lines by the final merged patch set. It capture the amount of all revision activities relative to the complexity of the merged patch set.
- $PC_{avg\_patchset}$ represents the average volume of edits (i.e., total number of added and deleted lines) across all patch sets of a code change. It is calculated by $PC_{revision} / (PC_{count} - 1)$.
- $PC_{max\_patchset}$ and $CC_{min\_patchset}$ indicate the largest and smallest patch set complexity, respectively, measured by the total number of added or deleted lines, found within any patch set of a given code change.

**Review Pattern (RP).** The RP features are designed to capture the interactions between an author and reviewer(s), such as patterns in code review discussions. Those features are to help us avoid the need for direct and complex semantic analysis of the review comments.

- $RP_{time}$ measures the time elapsed, in seconds, between the initial creation of a code change to its final submission.
- $RP_{weekday}$ indicates the day of week when a code change is submitted (e.g., starting from 1 for Sunday).
- $RP_{hour}$ indicates the hour of day when a code change is submitted. It uses a 24-hour format (e.g., 0 for [midnight, 1am)).
- $RP_{+2}$ is a boolean value indicating whether a code change is self- approved by the author. Self-approval occurs when the author gives a '+2' review score, while no other reviewer gives a positive score ('+1' or '+2').

The RP features primarily focus on the common review scenarios. Many RP variant features are not used in this paper due to either weak correlations with target result or their focus on uncommon scenarios. Here are the three examples. $RP_{review\_count}$ is for the total number of comments posted by all reviewers. $RP_{all\_clear}$ indicates unanimous positive reviews (e.g., 1 iff everyone commented gives a review score of '+1' or '+2'). It is to capture a case when a reviewer found a vulnerability but because only one engineer needs to give a '+2' score for submission, such valid concern was not properly addressed as part of the review. $RP_{last\_min\_change}$ captures if the author self-approved the final patch set with no positive review score on it. The difference with $RP_{+2}$ is whether some previous patch sets get a positive review score.

None of the RP features used in this paper is solely for rare cases. For example, let us consider a situation where a fixing plan is discussed in a code change (e.g., with multiple revisions), and then someone else cherry-picks the code change, amends it slightly, and self-approves for submission. While the $RP_{time}$ value of the cherry-pick change does not reflect the original review effort, the $RP_{+2}$ feature still captures the unusual case.

**Human History (HH).** The HH features aim to assess the creditability of individual engineers behind code changes. To this end, all past code change changes in the training dataset are classified as either likely-normal changes (LNCs) or vulnerability-inducing changes (ViCs). For each LNC, the author gets 2 points and every reviewer giving a review score of '+1' or '+2' gets 1 point. For each ViC, the author gets -3 points and a reviewer giving a positive review score gets -2 points. Those historical scores are then aggregated for each account. Using the aggregated data, the following HH feature values are calculated at runtime for every new patch set of any given code change:

- $HH_{author}$ measures the human history score of a code change author, calculated as a ratio of the ViC score of the author and the LNC score of the same author.
- $HH_{reviewer}$ measures the human history score of the reviewer(s) of a code change. Similar to the $HH_{author}$ score, the $HH_{reviewer}$ score of a reviewer is calculated as a ratio of the ViC score and the LNC score of the reviewer. For a code change with more than one reviewers, the highest value of all reviewers is used.
- $HH_{min\_reviewer}$ and $HH_{avg\_reviewer}$ are the minimum and average, respectively, human history score of all reviewers of a code change.

The quality of code changes by the same author (or reviewer) can fluctuate over time as engineers gains experience or falls behind on the latest vulnerabilities, secure coding practices, and product knowledge. To address this, the HH (Human History) features can be customized using sliding windows and weighted averages. That is to prioritize recent code changes and vulnerabilities over ones made a long time ago, ensuring the model adapts to the evolving skills and knowledge of engineers.

**Vulnerability History (VH).** The VH features aim to capture any patterns in when and where the vulnerabilities occurred. Those patterns encompass the temporal locality, spatial locality, and churn locality aspects [7]. Every change in the training dataset is classified as a LNC or ViC. Then the vulnerability history score of each file is calculated by (the number of LNCs in the file) $-3 \times$ (the number of ViCs in the file). Here, any file seen in the ViC list gets $-3$ points, while any file seen in the LNC list gets $+1$ point. Those VH statistics of files form the basis of the following VH features:

- $VH_{temporal\_max}$ and $VH_{temporal\_min}$ are the maximum and minimum, respectively, vulnerability history score value among all files within a given code change.
- $VH_{temporal\_avg}$ is the average of the same, reflecting the churn locality. As code changes involving many files often have relatively simple modifications per file, it takes an average value instead of an aggregated value.
- $VH_{spatial\_max}$, $VH_{spatial\_min}$, and $VH_{spatial\_avg}$ assess the spatial locality in the VH patterns. They consider the number of ViCs found in: (1) the files in the same directory as ones in a given code change, and (2) the files with the same file names (e.g., using different extensions) across all directories in the code change. The code change gets -2 points for every such a file. Similarly, the change gets +1 point for every LNC file in the same directory. The aggregated scores from all LNC files is then used as a denominator to normalize it (e.g., a code change with many files vs. few files).

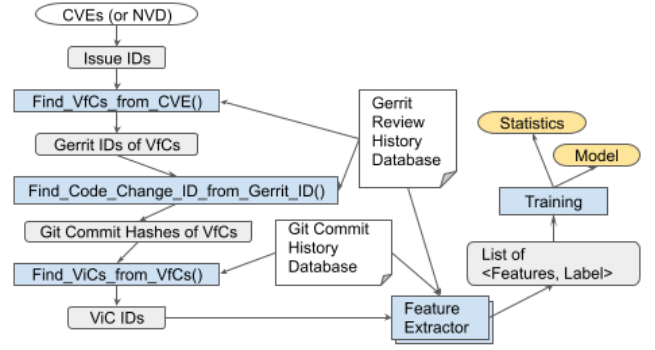

**FIGURE 2.** The end-to-end training pipeline of the presented system.

**Process Tracking (PT).** The PT features are designed to capture patterns in the volume of code changes throughout the software development lifecycle. It includes the trends in the numbers of LNCs (likely-normal changes), ViCs (vulnerability-inducing changes), and VfCs (vulnerability-fixing changes). For example, a mature project might see fewer (or less) code changes overall, fewer submitted ViCs, and a relatively increase in VfCs. Those trends are tracked by the following three PT features. Here, the trend values are pre-computed for each file in a target repository (e.g., all files in the entire git project).

- $PT_{change\_volume}$ measures the change in code volume (i.e., the number of source lines) between the current time period (e.g., a month) and the previous one.
- $PT_{VfC\_volume}$ measures the change in the number of VfCs submitted between the current and previous time periods. It is to assess the strength of underlying vulnerability triggers available in the target project. Its accuracy depends on how long it takes to discover vulnerabilities and then submit VfCs.

- $PT_{ViC\_volume}$ measures the change in the number of ViCs merged between the current and previous time periods.

**Text Mining (TM).** The TM features are extracted by analyzing the text content of code changes (specifically the added deltas) to identify semantic similarities between vulnerable code changes. The hypothesis is that vulnerable code changes would share common words or tokens in their source code or in accompanying code review discussions.

The TM features parse all the source code lines added by a given code change, identifying and counting specific code pattern types relevant to the target C/C++ programming language. The considered pattern types include: arithmetic (e.g., `+`, `−`, `*`, `/`, `%`), comparison (e.g., `==`, `!=`, `&&`), conditional (e.g., if, else, switch), loop (e.g., for, while), assignment (e.g., `=`, `<<=`, `+=`), logical (e.g., `&`, `|`, `^`, `~`), memory access (e.g., `->`, `.`), and all others. Before those text mining operations, all comments and string constants are removed during a tokenization process.

- $TM_{arithmetic}$ represents the proportion of arithmetic symbols found within the total count of all identified symbols.
- $TM_{comparison}$, $TM_{conditional}$, $TM_{loop}$, $TM_{assignment}$, $TM_{logical}$, and $TM_{memory\_access}$ are defined similar to $TM_{arithmetic}$.

While the above TM features are defined for source code deltas, similar metrics would be defined and applied to the textual content of review comments and code change descriptions, highlighting the flexibility of the TM features.

## V. DATA COLLECTION

This section describes how the vulnerability dataset is collected and generated for evaluating the accuracy of classifier models. The dataset consists of a list of source code changes where each change is labeled as ViC or LNC that includes VfC. Classification as either ViC or LNC aligns with the goal of this study to build a classifier that accurately differentiates between ViCs and LNCs.

The data collection process (depicted in Figure 2) involves the three key steps: (1) selecting all critical vulnerabilities found in the target AOSP codebase, (2) associating each vulnerability with its corresponding fixes, i.e., VfCs; and (3) locating of ViC(s) for each VfC:

**Selecting Target Vulnerabilities.** As depicted in Figure 2, this study leverages the CVE (Common Vulnerabilities and Exposures) database [7], maintained by the National Cybersecurity FFRDC (NCF), to select the target vulnerabilities. Specifically, it focuses on the CVEs that are found in the target AOSP codebase (namely, AOSP CVEs) and published on AOSP Security and Update Bulletins (ASB)[8].

This study excludes some types of CVEs to remain focused. First, self-discovered and fixed CVEs found internally by Google during new Android dessert releases (e.g., v14) are omitted due to the lack of publically available details. Second, vulnerabilities found in the proprietary extensions from silicon vendors, ODMs (e.g., Qualcomm), and the Google play service are not considered because they fall outside the upstream AOSP development. Third, CVEs of upstream Linux kernel (e.g., mainline, stable, and long-term releases) are excluded although AOSP-specific Linux kernel CVEs are included (e.g., ones found in the Android common kernel extensions). It is because they often involve code developed by Google, silicon vendors and ODMs, and are not strictly tied to a specific AOSP platform version.

**Associating Vulnerabilities and Fixes.** For each of the selected CVEs, this step locates the associated VfC(s). It begins by identifying all the relevant bug report(s) linked from a given CVE issue. We note that every target CVE issue published on the AOSP security bulletins has one or more associated bug reports stored in an issue tracking service (e.g., Google issue tracker [9] aka `Buganizer`). Conversely, multiple CVEs can sometimes share the same bug report if their fixes are identical or closely related.

Bug reports offer valuable insights into the vulnerability fixing process (e.g., key discussions done while reproducing or fixing them). In a vast majority of the cases, bug reports contain information about all or a subset of their VfCs. It is explicit if a VfC lists a bug report ID in its code change description (e.g., Bug: <number> or Fixes: <number> in the `gerrit` [10] change description) because then its submission event is posted on the bug report.

Our `BugID2GerritID` script automates the process of finding VfCs. It takes a list of bug IDs as input, scans the content of those bug reports, and returns any posted change IDs. Because code changes can be cherry-picked to other branches, a single change can exist across multiple branches. At this stage, the script does not yet differentiate between original changes and cherry-picks, gathering the change IDs (i.e., `gerrit` IDs) of all relevant changes.

While VfCs for CVEs or other important security issues usually reference their bug report in their `gerrit` description, in practice, depending on the used development protocol, it is not always the case. If the script finds no `gerrit` ID, a manual review work is triggered for all such bug reports to find the associated, implicit VfCs. Rarely, some bug reports do not have any VfCs if those externally known issues do not exist in the internal repository (e.g., already resolved).

Occasionally, such manual analyses relevel relevant `gerrit` changes or commits (e.g., URLs) linked to the VfCs. In those cases, the `GerritID2ChangeIDandCommitHash` script is used to extract the specific VfC IDs and commit hashes from the `gerrit` IDs. Importantly, commits sharing the same change ID indicate cherry-picks of the original change.

```python
01: def Find_ViCs_from_CVEs(CVEs):
02:   ViCs = {}
03:   for each CVE in CVEs:
04:     VfCs = Find_VfCs_from_CVE(CVE)
05:     ViCs.update(Find_ViCs_from_VfCs(VfCs))
06:   return ViCs
07:
08: def Find_ViCs_from_VfCs(VfCs):
09:   ViCs_dict = {}
10:   for each VfC in VfCs:
11:     ViCs = set()
12:     for each modified_file in VfC.files:
13:       for each modified_line in modified_file.lines:
14:         if IsEmpty(modified_line):
15:           continue
16:         if modified_line.type is 'delete':
17:           ViC.add(code change that added or
18:                   last modified modified_line)
20:         else:  # otherwise, it's 'add'.
21:           pass
22:           # Skips to group consecutively added lines
23:     ViCs_dict[VfC.id] = ViCs
24:   return ViCs_dict
```

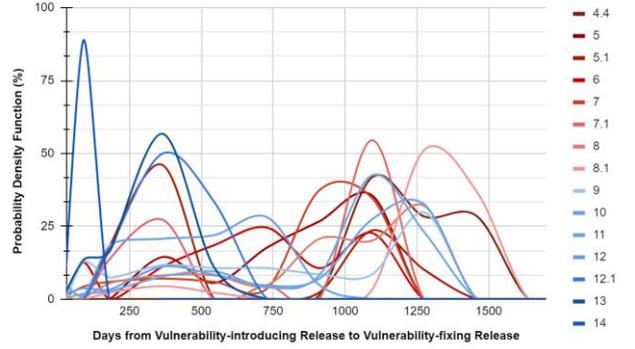

**FIGURE 3.** Probability density function for days from vulnerability-inducing release to vulnerability-fixing release of each AOSP release (in legend).

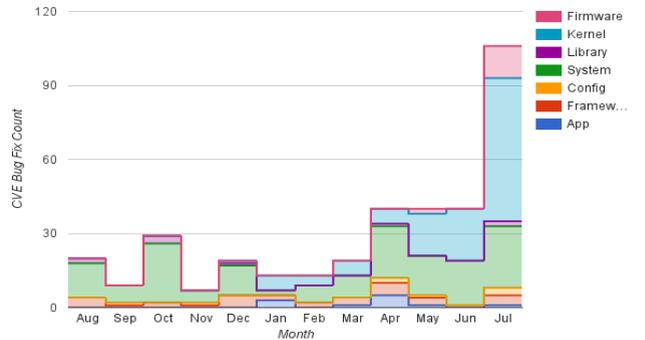

**FIGURE 4.** Monthly count of the AOSP CVE issues (using the first year data) grouped by the relevant system abstraction layer (in legend); Data for App and Framework layers is stacked at the bottom, following the order in the legend.

**Locating Vulnerability-inducing Changes (ViCs) for each Vulnerability-fixing Change (VfC).** The primary objective of Vulnerability Prevention (VP) is to maximize the accurate identification of ViCs. However, the aforementioned two steps in this section have thus far only allowed for the identification of CVEs and VfCs. Thus, we introduce a technique that enables the identification of ViCs from a given VfC. The identified ViCs undergo manual analysis to remove irrelevant code changes, resulting in a refined ViC set used to evaluate VP classifiers and features.

These identified ViCs Table I presents the algorithm for finding ViCs. It first identifies all the changed lines (i.e., additions and deletions) by using the git[11] show command and subsequently parsing its output data. For each of the identified, changed source code lines, our Blame script filters out extraneous lines (e.g., empty lines, headers, and comments) in order to only retain the relevant, vulnerability-fixing lines (VfLs). For a deleted line or a sequence of deleted lines, the script checks when each deleted line was added or last modified. The emphasis on the addition and last modification helps pinpoint potential ViCs because those code changes could have addressed the vulnerability at least but were unsuccessfully. We note that automatically and accurately determining whether a target vulnerability originates from the last modification or prior changes (if such changes exist) remains a challenge. Thus, this study relies on manual reviews for such cases.

When deleted lines are replaced by some newly added lines, typically more complex lines are added (e.g., in terms of the number of lines) to implement tailored error checking rules and error handling routines that can prevent a corresponding vulnerability at runtime. The tool does not classify such as a modification because it is challenging to determine whether it is a sequence of deletions and additions, or a true modification.

For an added line or consecutively added lines in VfLs, our Blame script analyzes when the next valid line was last modified. A next valid line, for example, means a line that is not an empty line nor a comment. It is to target the common case where an error checking routine is added right before a checked variable is used. By examining the addition or last modification time of the subsequent line, the tool identifies potential ViCs where the initial error checks for those variable(s) might have been missed.

If multiple ViCs are identified for a single VfC, the script lists them all. While the analysis done in this study mostly relies on such script-based automated techniques for locating ViCs, sometimes valuable insights for locating ViC(s) are found in the discussions posted on the bug reports or from the descriptions of VfCs. The tools and their algorithms are continuously refined through an iterative validation process of the discovered ViCs.

---

[11] Git is a distributed version control system used by AOSP. It is available at https://git-scm.com/.

## VI. CHARACTERIZATION

This section characterizes the collected vulnerability data. We note that Subsection VI.A utilizes all CVEs published in the Android Security Bulletins (ASB) from August 2015 to December 2023; Subsection VI.B utilizes CVEs published on ASB in the first year (from August 2015 to July 2016), and Subsection VI.C focuses on the CVEs found in the framework/av project of AOSP.

### A. VULNERABILITY FIXING LATENCY

Let us first analyze the time taken to detect and fix vulnerabilities in AOSP. Specifically, the number of days between each vulnerability-inducing release and its corresponding vulnerability-fixing release is measured. Figure 3 shows the results for each AOSP version (shown in the legend).

For a majority of the AOSP versions, *the measured vulnerability fixing latency peaks between 1,000 and 1,300 days (i.e., 3–4 years)*. The exception is seen in the recent releases (e.g., Android 13 and 14 released in <2 years) where the latency is also less than 2 years[12]. The tail is also long. For example, some vulnerabilities introduced in the two AOSP releases (e.g., v8.1) take over 4 years (>1,450 days) to be fixed.

While Figure 3 captures the time between vulnerability-inducing and fixing releases, it presents a conservative view. It excludes the time from the submission of a ViC to its corresponding AOSP release which is about a half year on average. Similarly, it does not include the time from a fixed AOSP release to OEM device updates [41][42]. Consequently, the true latency from ViC submissions to VfC rollouts to the user devices is longer (e.g., ~5 years instead of 4 years in Figure 3). Additionally, since the security update support window of the Pixel devices by an Android OEM is recently extended to 7 years in 2023 from the previous 5 years, the true latency for the older releases with the shorter support window could be longer than the data shown in Figure 3.

We note that the vulnerability fixing latency distribution is accurate for each AOSP dessert release version. However, it does not directly show the vulnerability fixing latency distribution of Android OEM devices in the field. It is because Android OEM devices are usually upgraded to newer Android dessert releases thanks to the fast software update efforts (e.g., TREBLE [37]) since Android 8.1. To show how to estimate the vulnerability fixing latency for OEM devices, let us consider an OEM device launched with Android 9.0, upgraded to Android 10 after one year, and upgraded to Android 11 after another year before reaching its End of Life (EoL). The vulnerability fixing latency for that OEM device can be calculated by concatenating: (1) the first year of vulnerability fixing latency data for Android 9.0;

TABLE II
PROJECTS WITH THE LARGEST NUMBER OF CVE FIXES

| Abstraction Layer | Project Path in AOSP | 1st Year Fixes |
|---|---|---|
| System | framework/av | 78 (47.3%) |
| | framework/native | 12 (7.3%) |
| | hardware/qcom/media | 9 (5.5%) |
| | external/libavc | 7 (4.2%) |
| | system/core | 6 (3.6%) |
| Kernel | drivers/staging/prima/CORE/HDD/src/ | 7 (6.9%) |
| | drivers/media/platform/msm/camera_v2/ | 6 (5.9%) |
| | drivers/media/platform/tegra/ | 6 (5.9%) |
| | fs/ | 6 (5.9%) |
| | drivers/misc/mediatek/com_soc/drv_wlan/mt_wifi/wlan/os/linux/ | 5 (4.9%) |
| | drivers/net/wireless/bcmdhd/ | 5 (4.9%) |
| | drivers/video/msm/ | 5 (4.9%) |
| | arch/arm/mach-msm/ | 4 (3.9%) |
| | drivers/video/tegra/host/ | 4 (3.9%) |
| | sound/soc/msm/qdsp6v2/ | 4 (3.9%) |

\* The ratios are relevant to all the fixes in their target abstraction layer.

(2) the first year of data for Android 10; and (3) the entire vulnerability fixing latency distribution for Android 11.

### B. ANALYSIS OF VULNERABILITY FIXING CHANGES

Given the observation that AOSP vulnerabilities can take over 4 years to fix, this analysis uses vulnerabilities fixed and published in the AOSP security bulletins during the first year (from August 2015 to July 2016). Those vulnerabilities are mostly found in the Android 4.1–6.0 releases, namely, Jelly Bean, KitKat, Lollipop, and Marshmallow.

**Vulnerability Fix Rate.** Over the analyzed one year, 356 CVEs are fixed, averaging approximately 0.975 ($\approx$ 1) CVE fixes per day. However, relatively large variations are seen in this rate across the 12 months as shown in Figure 4. It shows how the vulnerabilities fix pattern changed over the one year period. The CVE fix pattern shifts noticeably, with a sharp increase in the number of fixes during the final four months of the analyzed release period. *The surge aligns with approaching yearly AOSP and Pixel device releases.*

The seasonal pattern reflects the increasing focus on the security and stress testing as it gets close to the yearly release deadlines. Specifically, the emphasis during the initial months was on hardening the media and codec components of the Android native system. With the Android 7.0 (Nougat) release nearing, additional triggers were added to find the upstream Linux kernel vulnerabilities. Such shifts in testing focus are common during a software release lifecycle. Limited testing resources must be strategically allocated in accordance with development progress in order to ensure the quality, security, and other system integration requirements.

**Vulnerability Severity Distribution.** The severity data of the addressed CVEs reveals the importance of those fixes.

---

[12] Reference [49] reported ~1.98 years as the average vulnerability fixing latency using the initial Android security bulletins data (from August 2015 to November 2016). When that analysis was conducted, Android 6.0, 7.0,

and 7.1 were the most recent releases. It is consistent with our data (<2 years) for the current most recent releases (Android 13 and 14).

*About 82.9% (i.e., 32.9% critical and 50% high) of the fixed CVEs is categorized as critical or high.* Here, critical or high means the fixes are promptly created and integrated into the main and all the backport branches for monthly releases, expediting the fix rollouts compared to a annual update cycle from the main branch for the moderate or low severity issues. 15.7% of the same is classified as 'moderate' and only 1.4% is classified as 'low' or 'none'.

**Code Fixes for Vulnerabilities.** There is a many-to-many relationship between the CVEs and their code fixes.

- *1-to-1 relationship.* Typically, a single CVE issue fix is done by a single code change (e.g., git commit).
- *1-to-M relationship.* Some CVE fixes require multiple code changes. For the analysis purpose, code changes addressing the same CVE issue are grouped together if the changes are in a single git project. It reflects the observed common practice of developers splitting large fixes into smaller, more manageable code changes. Additionally, a code change related to deploying a fixed kernel image (e.g., to drop a rebuilt image to an Android repository) is considered part of the initial code change in a kernel code repository, as the change for a kernel image deployment stems from the initial source code change.
- *N-to-1 relationship.* Conversely, a single code change can sometimes resolve multiple CVEs. This is seen in cases of redundant CVEs for the same vulnerability or when multiple CVEs share a common root cause. Another example of an N-to-1 relationship is for when related CVEs exist for each affected device type (or chipset). Similar code changes applied to different device-specific branches are grouped together, including non-trivially cherry-picked changes with minor device- or chipset-specific adjustments. These semantically similar code changes are considered a single fix.
- *N-to-M relationship.* While it is rare, fixing CVEs with a seemingly N-to-1 relationship can sometimes involve more than one code changes. If distinct code changes remain across multiple system abstraction layers after the fore described grouping practices, the layer containing the most significant fix is prioritized for analysis.

**Abstraction Layers of Code Changes.** Figure 4 reveals the distribution of the first year AOSP CVE fixes across the system abstraction layers (or software subsystem-component types). Initially and consistently, many CVEs are addressed in the Android system layer[13] (such as the native servers, Hardware Abstract Layer modules, and Native Development Kit libraries [37]). Notably, the final quarter saw a significant increase in CVE fixes with the kernel layer.

Among the CVE fixes, nearly half (46.7%) target the Android system. A significant portion (33.5%) addresses the Linux kernel, while firmware fixes (such as bootloader) make up 4.2%. The remaining 19.8% is distributed as follows: Android app (~3.1%), Android Java framework (9.3%), other non-native code (5.1%), and configurations such as the SELinux policy, kernel config, init run command, and Android build rule (2.3%).

*The Android native software components are about 5.8 times more likely to contain the CVEs compared to the Android Java programs and configurations.* Table II shows the system and kernel projects with the most CVE fixes. The higher security of Java code stems from the two factors: the app store inspection process for Android apps and the inherent security benefits of type-safe Java and Kotlin programming languages used by Android apps and the Android framework. Here, the native software components are often developed by third-party contributors and other open source communities (e.g., GitHub and Linux kernel). However, vulnerabilities in the native code pose a significant security threat due to the powerful system privileges their attackers can exploit. Those low-level attacks can, in theory, subvert any overlying software running on top of the target layer and often do not require any user actions (e.g., app installation) to be triggered them. For example, it is possible to remotely exploit a system-level vulnerability through an MMS (Multimedia Messaging Service) message, even if the message is never opened by device users. As a result, it is often difficult to detect such system layer attacks.

Table II details the distribution of the system and kernel CVE fixes across their projects. The top five system projects in the table account for 67.9% of the first year, system-layer fixes. Notably, the framework/av project encompasses ~46% of the system-layer fixes, demonstrating the highest sample density. Within the kernel itself, drivers lead the pack with 72.5% of the CVE fixes, followed by the architecture-specific code (arch) at 8.8%, file system (fs) at 5.9%, and sound related code at 6.9%. The top 10 kernel projects encompass 51% of the first year, kernel CVE fixes.

## C. ANALYSIS OF VULNERABILITY-INDUCING CHANGES

For characterizing ViCs, let us focus on the CVEs fixed within the AOSP framework/av project, which exhibits the highest fixed vulnerability density. The project is a valuable target for in-depth ViC analysis due to the extensive testing (including fuzzing), the security hardening efforts in the Android Nougat release (e.g., vulnerabilities fixes), and its large size (e.g., 3,513 non-hidden files and directories, comprising 254,899 lines of C/C++ source code, configs, documents, and build rules).

---

[13] In [49] done using AOSP CVEs published between August 2015 and November 2016, 41% of CVEs are from the Linux kernel and 32% are from Android native libraries.

TABLE III
FEATURE PERFORMANCE USING AN N-FOLD VALIDATION (CLASSIFIER: DECISION TREE, N = 12)

| Metrics / Features | #LNCs classified as LNCs | #LNCs classified as ViC | #ViCs classified as LNC | #ViCs classified as ViC | LNCs Recall | LNCs Precision | ViCs Recall | ViCs Precision | ROC Area |
|---|---|---|---|---|---|---|---|---|---|
| HP | 7,453 | 0 | 585 | 0 | 1 | 0.927 | 0 | 0 | 0.541 |
| CC | 7,082 | 371 | 371 | 214 | 0.950 | 0.950 | 0.366 | 0.366 | 0.651 |
| RP | 7,012 | 441 | 361 | 224 | 0.941 | 0.951 | 0.383 | 0.337 | 0.666 |
| **HH** | **7,337** | **116** | **459** | **126** | **0.984** | **0.941** | **0.215** | **0.521** | **0.765** |
| **VH** | **7,275** | **178** | **270** | **315** | **0.976** | **0.964** | **0.538** | **0.639** | **0.833** |
| PT | 7,453 | 0 | 585 | 0 | 1 | 0.927 | 0 | 0 | 0.620 |
| TM | 7,223 | 230 | 442 | 143 | 0.969 | 0.942 | 0.244 | 0.383 | 0.588 |
| All | 7,188 | 265 | 230 | 355 | 0.969 | 0.964 | 0.573 | 0.607 | 0.786 |

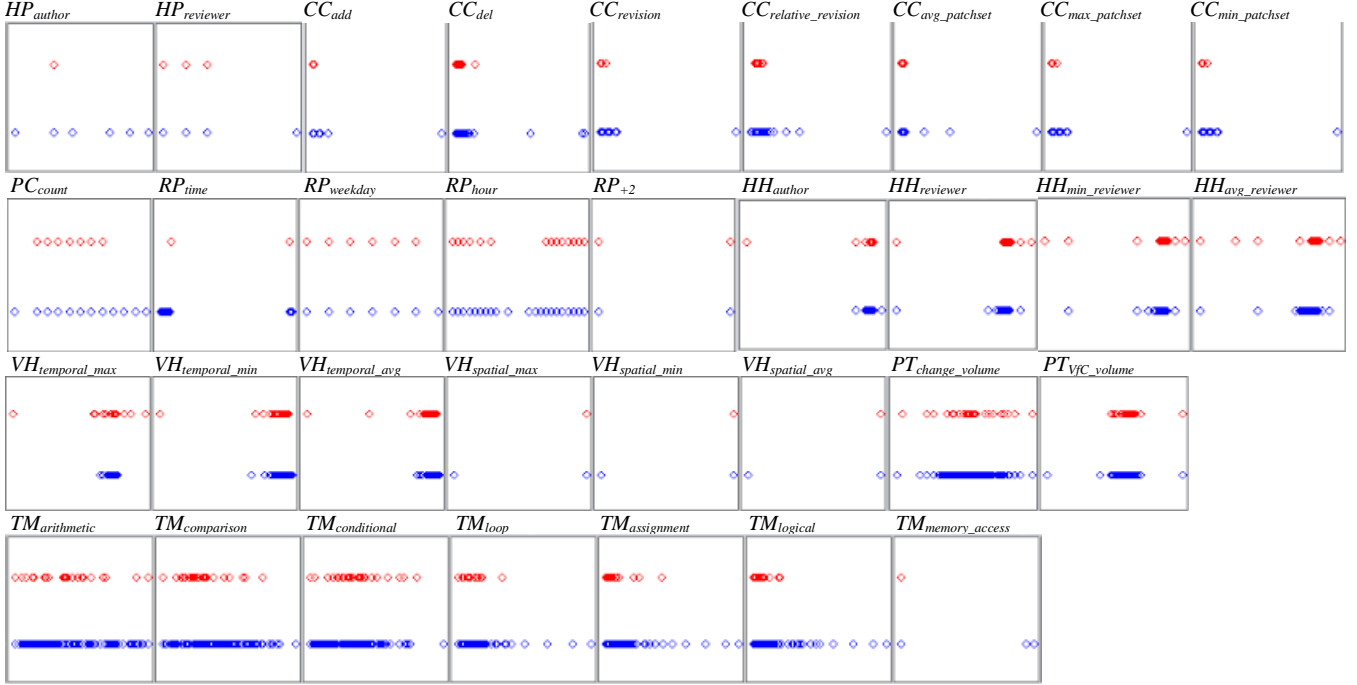

**FIGURE 5.** Visualization of values of each feature data type (x-axis is a value space, top red symbols are for ViCs, bottom blue symbols are for LNCs).

Fixing a single CVE issue can involve several VfCs. Of the 359 fixed CVEs analyzed, 77 require multiple VfCs that are merged into the target project. In total, those 77 CVEs are associated with 354 VfCs. Further analysis, using district code change identifiers, uncovers 244 unique VfCs. Our toolset then employs those unique VfCs to identify a total of 551 ViCs, which are subsequently characterized using our classification feature data types.

Table III summarizes the initial evaluation results for each feature set using a decision tree classifier. For example, the third column shows how many LNCs the VP framework predicts as ViCs. Notably, the HH (Human History) and VH (Vulnerability History) feature sets achieve high accuracy in ViC identification. Conversely, neither the HP (Human Profile) nor PP feature sets detect any ViCs, while the remaining feature sets exhibit varying accuracy levels.

Figure 5 visually analyzes feature values to provide deeper insights into the effectiveness of different feature sets. It shows the distribution of feature values for both ViCs (red symbols, upper row) and LNCs (blue symbols, lower row). The x-axis represents the value range of each specific feature data type. The visualization reveals patterns explaining why certain feature sets perform better than others in predicting ViCs.

The HP feature set shows limited effectiveness in AOSP because it relies on two discrete features. ViCs tend to cluster within a narrower range of those feature values compared to LNCs (e.g., ViCs utilize only one value of the $HP_{author}$ feature). The limited value distribution likely stems from the target project development being primarily handled by a single organization, fostering consistent coding practices within the AOSP framework codebase. Consequently, *code change author affiliation is not a strong predictor of vulnerabilities within AOSP*.

The initial hypothesis that malicious external contributors were a primary source of vulnerabilities proves incorrect in Android platform developments. The data analysis reveals that most ViC authors are not malicious third-party actors. It

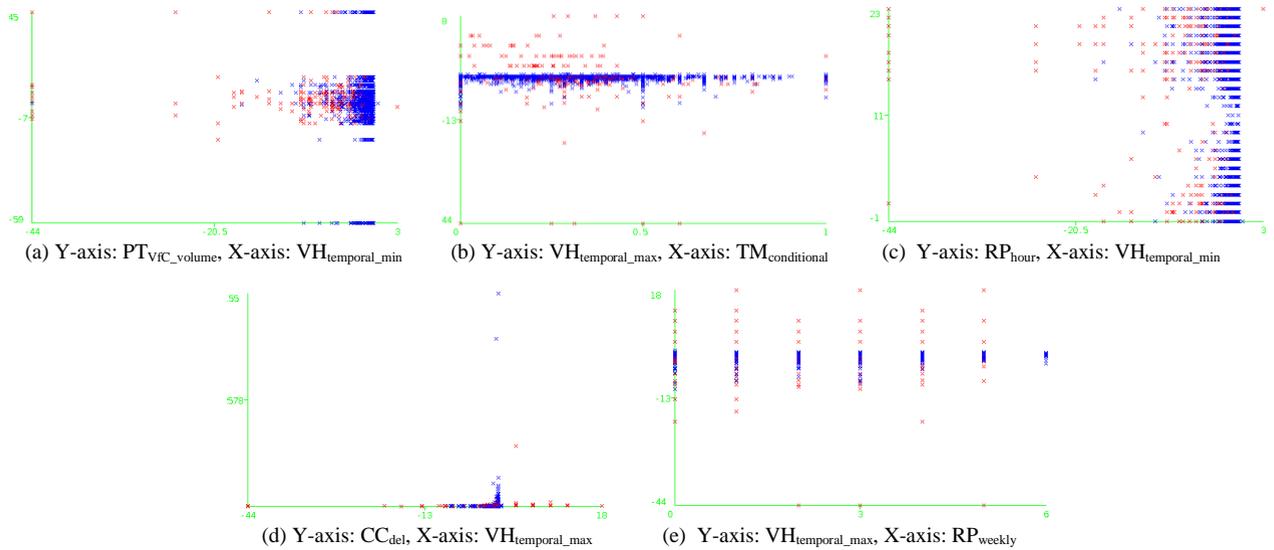

(a) Y-axis: $PT_{VfC\_volume}$, X-axis: $VH_{temporal\_min}$

(b) Y-axis: $VH_{temporal\_max}$, X-axis: $TM_{conditional}$

(c) Y-axis: $RP_{hour}$, X-axis: $VH_{temporal\_min}$

(d) Y-axis: $CC_{del}$, X-axis: $VH_{temporal\_max}$

(e) Y-axis: $VH_{temporal\_max}$, X-axis: $RP_{weekly}$

**FIGURE 6.** Multi-variant analysis examples using a pair of features where red symbols are for ViCs and blue symbols are for LNCs.

is likely due to the rigorous collaboration process in place for external contributions to AOSP: such contributors usually lack direct commit permissions, and their code changes can sometimes undergo extensive scrutiny by the project owners. Thus, the observation is *AOSP vulnerabilities are more likely to arise when both authors and reviewers are trusted entities and consequently there is reduced inspection and testing thoroughness.*

The CC (Change Complexity) feature set reveals a pattern. Most ViCs involve small- or medium-sized code changes, while LNCs exhibit a wider range of sizes, encompassing both tiny and extra-large code changes. It suggests that *code modifications exceeding a certain size threshold (e.g., >250 lines) would introduce enough complexity to distract both authors and reviewers, increasing the likelihood of undetected vulnerabilities.* However, some extremely large code changes often involve repetitive or mechanical edits (e.g., pattern-based refactoring or removing deprecated code) rather than modifications to intricate logic, making them less prone to oversights. Interestingly, the $CC_{revision}$ feature also indicates that *ViCs typically undergo fewer revisions during their code reviews compared to LNCs.* The observation supports the idea that some LNCs may initially contain vulnerabilities that are addressed through the code review process, leading to more revisions.

The HH (Human History) feature set confirms a trend. In general, *authors and reviewers previously involved in ViCs are more likely to be associated with the introduction of new ViCs.* This pattern is evident in Figures 5 ($HH_{author}$ and $HH_{reviewer}$ sub-graphs), where ViCs exhibit high-density clusters slightly to the right of LNC value clusters. The sparse distribution on the left side of the upper row (representing individuals with only one ViC at the time of analysis) is likely to converge towards the right side cluster for ViCs over time. This finding highlights *the importance of identifying ViCs and providing early, targeted feedback to the involved software engineers.* Such feedback can improve their understanding of vulnerabilities, aiding prevention efforts in the near future.

The VH (Vulnerability History) feature set indicates ViCs and LNCs generally modify a similar set of files. However, some ViCs introduce changes to previously untouched files. *Such modifications on untouched files consistently result in vulnerabilities in the analyzed dataset.* It can be explained by the two scenarios: a newly created file is modified for the first time, introducing a ViC, or a file undergoes multiple local edits that are later combined (e.g., using `git squash` mechanism) into a commit (ViC) visible on the main repository. The practice of infrequently upstreaming large, merged changes potentially increases the risk of vulnerabilities.

This paper prioritizes characterization of impactful feature sets. Other individual feature sets are omitted due to redundancy with the fore described characteristics or the lack of clear patterns in their visualizations in Figure 5. It, at the same time, underscores the importance of multivariate analysis, as demonstrated in Figure 6. Here, specific combinations of two features (i.e., 5 pairs in total) yield relatively effective classifiers with clear clustering patterns (or hyperplanes) in the two-dimensional space. Analyzing only single features or pairs would provide an incomplete understanding of the true potential of the entire feature sets, given the numerous informative combinations possible. Thus, a comprehensive evaluation study is crucial.

## VII. RESULT

The section conducts a comprehensive evaluation of the accuracy of our framework across both the training and inference phases, reflecting real-world performance.

| Classifiers \ Metrics | #LNCs classified as LNC | #LNCs classified as ViC | #ViCs classified as LNC | #ViCs classified as ViC | LNCs | | ViCs | | ROC Area |
|---|---|---|---|---|---|---|---|---|---|
| | | | | | Recall | Precision | Recall | Precision | |
| **Random Forrest** | **7391** | **62** | **233** | **352** | **0.992** | **0.969** | **0.602** | **0.850** | **0.955** |
| Decision Tree | 7188 | 265 | 230 | 355 | 0.964 | 0.969 | 0.607 | 0.573 | 0.786 |
| Quinlan C4.5 | 7270 | 183 | 255 | 330 | 0.975 | 0.966 | 0.564 | 0.643 | 0.832 |
| Logistic Regression | 7380 | 73 | 343 | 242 | 0.990 | 0.956 | 0.414 | 0.768 | 0.918 |
| Naïve Bayes | 7278 | 175 | 353 | 232 | 0.977 | 0.954 | 0.397 | 0.570 | 0.866 |
| SVM (SMO) | 7444 | 9 | 476 | 109 | 0.999 | 0.940 | 0.186 | 0.924 | 0.593 |

### A. N-FOLD VALIDATION

We first identify the optimal classifier type, followed by the feature dataset reduction.

**Classifier Selection.** To select the most accurate classifier type, all six types of classifiers are evaluated using the complete set of devised feature data. The training dataset incorporates information about all known ViCs. The evaluation employs the Weka v1.8 [38] toolkit with the default parameter configurations for each classifier, ensuring a fair comparison of their inherent performance.

Table IV shows the 12-fold validation result. The Random Forest classifier demonstrates the highest classification accuracy among the six types tested. It achieves ~60% recall for ViCs with 85% precision, while misclassifying only 3.9% of LNCs (calculated as 1–0.992×0.969). *Based on the evaluation result, the rest of this study uses Random Forrest.*

The superior performance of Random Forrest over the Decision Tree classifier is expected, as shown by the relative operating characteristic curve (ROC) area of 0.955 vs. 0.786. The Quinlan C4.5 classifier also maintains the notably lower precision and recall than Random Forest for classifying ViCs.

The logistic regression classifier exhibits the second-best performance in terms of ROC area, mainly thanks to its relatively high precision (0.768) for ViCs. However, its recall for ViCs is significantly lower (0.414) compared to the Decision Tree and Quinlan C4.5 classifiers. Similarly, the naïve Bayes classifier underperforms the logistic regression classifier across all three metrics (recall, precision, and ROC area).

Finally, the SVM classifier demonstrates the highest recall for LNCs and a good precision for LNCs, indicating that the model is over-fitted to the LNC samples. It can be confirmed by the fact that SVM does not show a good recall for ViCs. The over-fitting is likely because of the imbalanced training dataset, where the LNC samples significantly outnumber the ViC samples. SVM performance generally benefits from a balanced ratio of positive and negative examples (e.g., 1:1), which is particularly difficult in vulnerability classification tasks.

**Feature Reduction.** Let us evaluate the performance of the Random Forest classifier using various subsets of the devised feature data types. The process is to identify a highly effective feature subset that maintains high accuracy, while requiring less data collection during inference compared to using the full feature datasets.

Table V presents the evaluation results. As expected, the first row, using all six feature sets (VH, CC, RP, TM, HH, and PT) represents the best case. Removing the HH (Human History), PT (Process Tracking), or TM (Text Mining) feature sets individually leads to minor reductions in the recall (0.4–0.9%) and precision (0.6-2.8%) for classifying ViCs. Practically, it translates to ~5 misclassified ViCs out of the 585 ViCs and ~14 misclassified LNCs out of the 7,453 LNCs. The ROC area remains largely consistent across those three variations (0.954–0.957 for the 2nd, 3rd, and 4th rows), compared to the baseline of 0.955 (the 1st row in Table V).

Let us further investigate the accuracy achieved after removing both the HH (Human History) and PT (Process Tracking) feature sets, followed by the removal of all three (HH, PT, and TM). The results show that the VH (Vulnerability History), CC, and RP (Review Pattern) feature sets still provide high accuracy, exhibiting only a 0.3% reduction in LNC recall and a 4.5% reduction in ViC precision over when all features are used. The following discusses each of the three remaining feature sets in more details:

The VH (Vulnerability History) feature set aligns with the known factors used in the buggy component prediction (e.g., temporal, spatial, and churn localities). The results in this study demonstrate that those three types of *localities remain relevant and effective for predicting vulnerabilities at the code change level.* Among the six VH feature data types, $VH_{temporal\_avg}$ *is the most impactful.* It is confirmed by the fact that none of the other five VH feature data types alone could correctly classify a single ViC in isolation during the 12-fold validation experiment.

The CC (Change Complexity) feature set aligns with the established principle that complexity often leads to software defects, a relationship repeatedly observed when analyzing defect ratios of software components (e.g., files or modules). The data in this study further confirms that *more complex code changes are indeed more likely to introduce vulnerabilities.* Our VP framework thus signals software engineers to pay extra attention by selectively flagging a subset of code changes as higher risk (e.g., using predicted chances of vulnerabilities) It is to help identify and fix potential coding errors before those code changes are merged into a source code repository.

| Features | #LNCs classified as LNC | #LNCs classified as ViC | #ViCs classified as LNC | #ViCs classified as ViC | LNCs Recall | LNCs Precision | ViCs Recall | ViCs Precision | ROC Area |
|---|---|---|---|---|---|---|---|---|---|
| VH+CC+RP+TM+HH+PT | **7391** | 62 | 233 | **352** | 0.992 | 0.969 | 0.602 | 0.850 | 0.955 |
| VH+CC+RP+HH+PT | *7380* | *73* | *237* | *348* | *0.990* | *0.969* | *0.595* | *0.827* | *0.957* |
| VH+CC+RP+TM+HH | **7389** | 64 | 238 | **347** | 0.991 | 0.969 | 0.593 | 0.844 | 0.957 |
| VH+CC+RP+TM+PT | **7377** | 76 | 235 | 350 | 0.990 | 0.969 | 0.598 | 0.822 | 0.954 |
| VH+CC+RP+TM | 7371 | 82 | 239 | 346 | 0.989 | 0.969 | 0.591 | 0.808 | 0.953 |
| VH+CC+RP (*) | **7368** | 85 | 233 | 352 | **0.989** | **0.969** | **0.602** | **0.805** | **0.952** |
| VH+CC | 7200 | 253 | 221 | 364 | 0.966 | 0.970 | 0.622 | 0.590 | 0.799 |
| CC+RP | **7429** | 24 | 399 | 186 | **0.997** | **0.949** | 0.318 | **0.886** | **0.802** |
| RP | 7320 | 133 | 390 | 195 | 0.982 | 0.949 | 0.333 | 0.595 | 0.720 |
| VH | **7330** | 123 | **270** | 315 | **0.983** | **0.964** | **0.538** | **0.719** | **0.918** |

| Features | #LNCs classified as LNC | #LNCs classified as ViC | #ViCs classified as LNC | #ViCs classified as ViC | LNCs Recall | LNCs Precision | ViCs Recall | ViCs Precision | ROC Area |
|---|---|---|---|---|---|---|---|---|---|
| VH+CC+RP+TM+HH+PT | 7,391 | 62 | 233 | 352 | 0.992 | 0.969 | 0.602 | 0.850 | 0.955 |
| CC+RP+TM+PT | 7,444 | 9 | 398 | 187 | ***0.999*** | 0.949 | *0.320* | ***0.954*** | 0.829 |
| CC+RP+PT | 7,432 | 21 | 398 | 187 | ***0.997*** | 0.949 | *0.320* | ***0.899*** | 0.818 |
| CC+RP | 7,429 | 24 | 399 | 186 | ***0.997*** | 0.949 | *0.318* | ***0.886*** | 0.802 |
| $CC_{add}+CC_{revision}+CC_{relative\_revision}+CC_{avg\_patchset}$ $+PC_{count}+RP_{time}+RP_{weekday}$ | 7,421 | 32 | 399 | 186 | ***0.996*** | 0.949 | *0.318* | ***0.853*** | 0.801 |
| $CC_{add}+CC_{revision}+CC_{relative\_revision}$ $+RP_{time}+RP_{weekday}$ | 7,385 | 68 | 397 | 188 | ***0.991*** | 0.949 | *0.321* | ***0.734*** | 0.786 |

The data confirms the importance of the novel RP (Review Pattern) feature set in the VP framework. Complex code changes are likely to contain software faults, placing a burden on code reviewers to detect coding errors and guide authors toward fixes. While the *RP feature set alone does not provide the highest accuracy for ViCs (e.g., 59.5% precision), combining it with the CC (Change Complexity) feature set significantly boosts the precision for ViCs (e.g., 88.6%).* The pairing helps identify situations such as: when complex code changes lack rigorous review before submission; or when authors self-approve complex changes without any explicit peer code reviews recorded. However, *the RP and CC feature sets do not offer high ViC recall* (e.g., 31.8%) as the pair target the specific code change characteristics. Many other factors contribute to ViCs slipped through code reviews and other pre-submit testing.

Further removing RP (Review Pattern) from the VH, CC, and RP sets significantly reduces accuracy (i.e., 80.5% precision for ViCs drops to 59%). Interestingly, even when both RP and CC (Change Complexity) features set are removed, the VH (Vulnerability History) features set alone still provides the higher accuracy than the VH and CC sets combined (e.g., the ROC area of 91.8% vs. 80.2%). It is partly due to VH leveraging the N-fold validation setting (i.e., learning from future ViCs to predict past ViCs). The next subsection (VII.B) addresses it using online inference and demonstrates a general counterexample applicable to all feature data types.

**Potential as a Global Model.** To enable immediate deployment of a VP model across multiple projects, this study also investigates which feature data types are likely target project agnostic. Among the six feature sets, four (CC,

RP, TM, and PT) are potentially not project-specific. In contrast, the HH (Human History) and VH (Vulnerability History) feature sets focus on vulnerability statistics tied to specific engineers and software modules, respectively. It suggests us that models trained using those two feature sets would not be directly transferable to other projects with different engineers and software modules.

We explore the possibility of a global VP model though another 12-fold validation study. Because one may argue TM (Text Mining) could be programming language-specific, the accuracy of a global model is evaluated without and with TM to assess its impact. Table VI shows that using only the CC, RP (Review Pattern), and PT (Process Tracking) feature sets yields relatively low ViC recall (32%) but notably high ViC precision (~90%). The precision increases further if TM is used together with CC, RP, and PT (~95%). The result is promising, as those feature sets could potentially be used across multiple projects due to their ability to minimize false positives.

Let us further reduce the feature sets by considering individual features. Using only the five features listed in the last row of Table VI ($CC_{add}$, $CC_{revision}$, $CC_{relative\_revision}$, $RP_{time}$, and $RP_{weekday}$) the VP model achieves a ROC area of 0.786, while maintaining a high ViC precision of 73.4%. It comes at the cost of a notable reduction in recall (i.e., to ~32% from ~60% when all feature sets are used). However, we argue that the penalty is minimal, as evidenced by the still-high LNC precision of 94.9%. Importantly, our approach remains significantly better than not using the VP framework at all, since it retains a ViC recall of 32.1%.

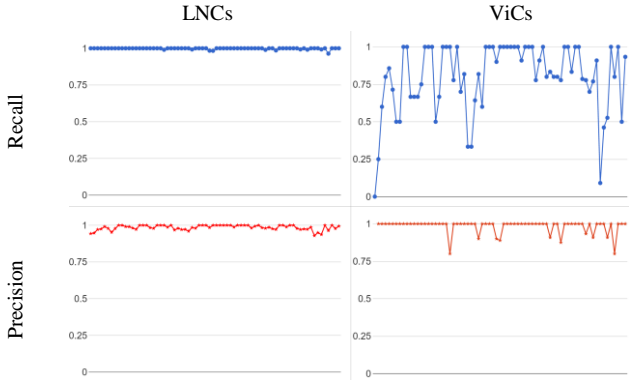

**FIGURE 7.** Online deployment mode result; x-axis is in month (~6 years total).

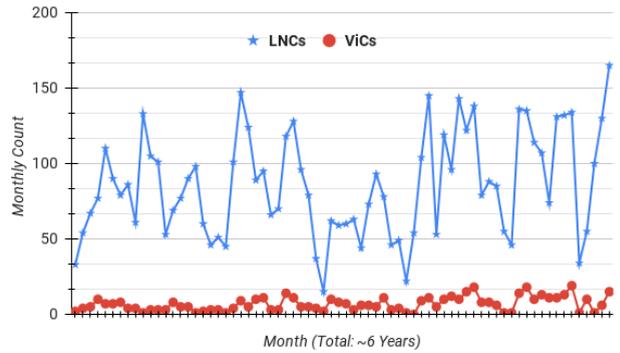

**FIGURE 8.** Monthly Counts of LNCs and ViCs.

In our N-fold cross-validation, the recall for ViCs is not notably high. Yet it confirms the extra security coverage that can be provided by the VP framework without having to conduct extensive security testing. The relatively low recall is likely due to the validation process not fully capturing the inherent temporal relationships, dependencies, and patterns within the feature data. For instance, N-fold validation can reorder a ViC and its corresponding VfC such that the VfC precedes the ViC, violating the natural order. Consequently, we evaluate the VP framework using its online deployment mode to better reflect real-world scenarios.

### B. EVALUATION USING ONLINE DEPLOYMENT MODE

This subsection evaluates the VP framework under its production deployment settings (namely, online deployment mode) using about six years of AOSP vulnerabilities data.

To achieve maximum accuracy, this experiment employs the Random Forest classifier and leverages all devised feature data types. The evaluation data originates from the AOSP frameworks/av project. Each month, the VP framework assesses all code changes submitted in that month using the latest model, trained on data available before that month begins. For this evaluation, it is assumed that a ViC is known if and only if it is merged. However, a more realistic scenario considers a ViC known if its corresponding VfC is merged. The assumption highlights the need for thorough security testing (e.g., fuzzing) to identify ViCs within an average of half a month after they are merged. Thus, *existing security testing techniques are crucial to fully realize the potential of the VP framework.*

Figure 7 presents the evaluation results of the online deployment mode. *For ViCs, the framework demonstrates an average recall of 79.7% and an average precision of 98.2%. For LNCs, it achieves an average recall of 99.8% and an average precision of 98.5%.* Those results indicates that the online VP framework can identify ~80% of ViCs with ~98% accuracy, while only misdiagnosing ~1.7% of LNCs (assuming no hidden vulnerabilities within LNCs) at pre-submit time before code changes are merged. The actual misdiagnosis rate is likely lower than 1.7% due to potential future discovery of vulnerabilities within LNCs. Similarly, the exact ViC accuracy metric values can change depending on the classification of newly discovered ViCs in the future. The direction and magnitude of such metric value changes depend on how those new ViCs were previously classified. Overall, these promising results warrant further investigation for industry and open source community deployments.

Significant variations exist in the ViC recall values (i.e., a standard deviation of 0.249). While one might assume low recall in months with few ViCs (e.g., <5), the sample correlation coefficient analysis shows no significant link (-0.068) between the ViC recall and count (captured in Figure 8). In contrast, the LNC recall and precision values show less variations. Among those two, the precision exhibits slightly wider variation than the recall (i.e., a standard deviation of 0.017. It likely stems from the abundance of LNCs each month and the high LNC precision of the VP framework (as shown in Subsection VI.A).

The online mode demonstrates notably higher accuracy than the 12-fold validation using the same feature data and classifier. It is likely due to the online mode giving greater weights to recent history within its learning model, effectively leveraging the strong temporal correlations found in certain feature values. For example, a file in a ViC is likely to contain another ViC in the near future if the same software engineers continue working on the file (e.g., as author and reviewers) and are performing similar tasks (e.g., as part of a workstream to develop a new feature). The 12-fold validation, with its shuffled training and test data, does not fully capture such temporal causality. Consequently, the online mode results provide a more realistic assessment of the VP framework accuracy than the 12-fold validation ones.

Figure 8 reveals that *an average of 7.4% of reviewed and merged code changes are classified as ViCs. The framework flags an average of 6.875 LNCs per month for additional security review. This manageable volume (<2 code changes per week) represents an acceptable review cost*, especially considering the large number of full-time software engineers worked on the target project.

### VIII. DISCUSSION

This section discusses the implications for multi-project use and the previous Android security works; threats to validity; and alternative approaches.

### A. IMPLICATIONS ON MULTI-PROJECTS

This subsection explores adapting the VP framework for multi-project use cases. There are two options for model training. One is to train a separate model on data from each project and apply it individually. The other is to identify project-agnostic features to build a global model applicable across multiple projects. The latter option differs from [5], which explored applying the training data of a single project to different projects. Since the latter global model option is described in Subsection VII.A, this subsection focuses on discussing the former option.

**Project-Specific Model.** The VP framework is adaptable to any open-source project with tracked vulnerability issues (e.g., CVEs). For a target project, the framework follows a sequence of: (1) identifying vulnerability issues, corresponding VfCs, and ViCs; (2) deriving a feature dataset from the identified data; and (3) training a classifier model, which is continuously updated as new vulnerabilities are discovered.

While using the VP framework itself does not require a high density of ViCs, effective evaluation of the framework against a target project does. A project with many discovered CVEs indicates a high density of ViCs. Thus, the experiment in Subsection VI.B targets the AOSP frameworks/av project, known for its abundance of CVEs (e.g., over 350). It is because extensive fuzzing and other security testing were conducted on that project (e.g., between 2015 and early 2016 prior to the Android Nougat release). As fuzz testing expands to other Android components (e.g., native servers and hardware abstraction modules), we believe that future studies can leverage our experiment method in order to more comprehensively evaluate the accuracy of the VP framework or a new variant technique across a wider range of AOSP projects.

### B. IMPLICATIONS ON ANDROID SECURITY WORKS

Although it is not the primary theme of this paper, another key contribution is the definition and measurement of vulnerability fixing latency distributions per major Android release. The measurement of vulnerability fixing latency has two implications:

**Vulnerability-fixing latency is longer than software update latency in the Android device ecosystem.** Our metric focuses on capturing the time it takes to detect and fix vulnerabilities. Figure 9 illustrates the overall Android software update process, beginning when a vulnerability-inducing change (ViC) is merged into a repository. While some ViCs are detected (i.e., as issues in Figure 9) and fixed internally, others remain undetected and are included in AOSP releases, which are then shipped to end-user devices through OEMs and carriers. After the initial release, those vulnerabilities are discovered and fixed, and then published as CVEs and on AOSP security bulletins. Our vulnerability fixing latency metric measures the time from the ViC merge to the publication of fixes (i.e., VfC in Figure 9) and CVEs on AOSP security bulletins.

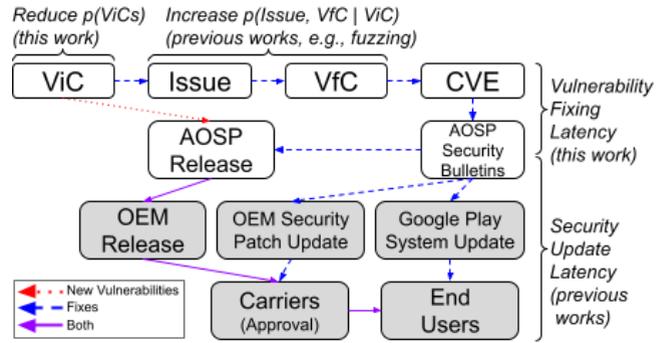

**FIGURE 9.** Android software update process for vulnerabilities and fixes.

A significant portion of prior work focuses on reducing and measuring the time to update in-market end-user devices with available fixes. Existing studies on security updates [41][42][43][44][50] primarily measure the time from AOSP security bulletins to end-user OTA (over the air) updates. That security update time is considerably shorter than the vulnerability fixing latency reported in this study. For example, [42] indicates that updates can take ~24 days for devices with monthly security updates [41] and 41-63 days for Android OEM devices with quarterly or biannual updates when considering the update frequency, device model, and regional factors. Given that the software update time is relatively short compared to vulnerability fixing latency, our data suggests *the Android security research community should focus more on reducing vulnerability fixing latency to optimize the end-to-end latency from vulnerability induction to end-user device updates.*

**Efforts to prevent vulnerabilities are justified given the difficulty of identifying and fixing vulnerabilities after introduction.** In addition to measurement, existing Android security work focuses on accelerating both: (1) the software update process (such as TREBLE and Mainline[14]) and (2) the vulnerability identification process. TREBLE modularizes the Android platform, allowing for fast updates of the silicon vendor-independent framework by OEMs. Mainline enables updates to the key Android system-level services and modules via Google Play Store system app updates (see Google Play System Update in Figure 9). Extensive fuzzing and other security testing efforts are employed to increase vulnerability detection and accelerate fixes [45][46][47][48]. *Those previous works aim to increase the probability of*

---

[14] Android 10 release notes, available at
https://source.android.com/docs/whatsnew/android-10-release

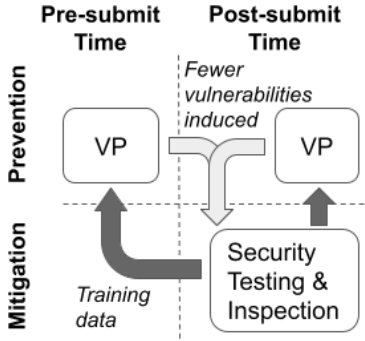

**FIGURE 10.** Synergies between the presented VP approach and the existing security testing and inspection techniques.

*identification and fixes given a vulnerability-inducing change, P(Issues, Fixes | ViC), as shown in Figure 9. However, they do not directly reduce P(ViC) itself.* The vast AOSP codebase and high frequency of code changes across various software abstraction layers for different hardware components and devices make it difficult to keep up with new vulnerabilities – as captured by the vulnerability fixing latency metric.

This paper thus introduces the concept of pre-submit, Vulnerability Prevention (VP) at the upstream software project level, as an alternative approach. VP aims to directly reduce *P(ViC)* – the probability of having vulnerability-inducing changes. We demonstrate that our VP approach could have been effectively used to protect a key AOSP project from the historical ViCs. The cost of using VP is also relatively small (e.g., in terms of computing power) compared with some existing security testing techniques, such as fuzzing and dynamic analysis.

### C. THREATS TO VALIDITY
This subsection analyzes potential threats to the validity of this study, including a discussion of underlying assumptions.

The N-fold validation experiments in this study assume timely discovery of ViCs (vulnerability-inducing changes). It highlights an iterative process depicted in Figure 10. That is, the improved knowledge of ViCs enhances the VP accuracy, which then aids in the ViC prevention or discovery in both upstream (pre- or post-submit) and downstream projects (e.g., before integrating upstream patches or releasing new builds). It implies that ViCs prevented by our VP framework (e.g., with revisions made in response to flagging) can be incorporated into model training to further reinforce its accuracy.

To bootstrap our VP process, initial extensive fuzzing is crucial. In practice such preventive security testing usually targets software components known to be vulnerability-prone – a decision guided by security engineering considerations. This study also uses the `frameworks/av` component as it meets this criterion. Alternatively, one could start with our global VP model for high precision detection of a subset of ViCs. The resulting initial ViC dataset can then be used to train project-specific VP models by leveraging a wider range of feature data types than the global model.

This study implicitly assumes a focus on C/C++ projects, which are a major source of AOSP vulnerabilities. Java would have slightly different characteristics, e.g., affecting the TM (Text Mining) feature set, potentially requiring customization of effective features. This study also assumes an AOSP-like development model with registered partners (e.g., employees from public companies) as primary contributors. We recognize that volunteer-driven open-source projects (e.g., with fewer, non-obligated developers and reviewers) would present different dynamics. Furthermore, the commercial use of Android on billions of devices drives a strong focus on code quality among its developers. The commitment to quality is evident in many other upstream and downstream open-source projects, although often to a lesser degree. Thus, while optimal feature sets vary across open source projects, the data in this study shows us that the core VP approach (consisting of the framework, tools, feature types, training process, and inference mode) remains broadly applicable within the upstream open source domain, understanding the high cost of fixing vulnerabilities merged downstream.

The accuracy of results depends also on the algorithm used to identify ViCs from VfCs (vulnerability-fixing changes). The manual analysis of random samples used to refine our algorithm generally showed that identified ViCs, at minimum, touch code lines near where the respective vulnerabilities are introduced. While it can be difficult to definitively link a code change to a vulnerability (or other semantically-sophisticated defects), changes made in close proximity to the vulnerability-inducing lines are more likely to be noticed during code review than those made farther away.

### D. ALTERNATIVE APPROACHES
This study also considered some alternative approaches:

Using the HH (Human History) features, one can identify a subset of software engineers who have reviewed a sufficient number of code changes without missing a significant number of ViCs. When the VP framework flags a code change, it can then be assigned to one of those trusted reviewers for approval. It offers an alternative way to establish a secure code reviewer pool, especially when a dedicated internal security expert group is unavailable.

Requiring a security test case for every fixed CVE issue offers clear benefits (e.g., to confirm corresponding VfC and detect regressions). However, it is important to acknowledge the practical limitations. Recent AOSP CVE fixes often come with the proof-of-concept (PoC) exploits that developers can leverage to reproduce and validate their fixes. Those PoC security tests can either exploit vulnerabilities or confirm the mitigation effectiveness when that is feasible with high success probability. In practice, we recognize that mandating security tests for every CVE (or VfC) is resource-intensive and cannot be universally adopted across projects and organizations due to variations in engineering cultures and processes.

While it is possible to write and run a comprehensive set of security test cases for pinpointing ViCs across historical commits, this alternative approach has inherent limitations. Using a bisecting algorithm to identify the first or last ViC still incurs significant costs compared to our ML-based approach. It is particularly true because vulnerabilities often manifest only on specific devices. It necessitates testing on a wide range of devices unless affected devices can be precisely identified. However, access to such devices is often limited. Furthermore, the reliability of test results is compromised if the security tests exhibit flakiness (i.e., sometimes failing even when they should be passing).

## IX. RELATED WORK

This section reviews the related works.

**Main Focus of System Security Research.** Historically, much of the security engineering effort was focused on identifying previously unknown types of vulnerabilities and designing their mitigation techniques. Consequently, system security research often emphasized the development of innovative attack and defense techniques [39][40]. To this end, such previous security works employed a range of security testing, validation, and verification techniques [45][46][47][48], including stress testing, instrumentation, fuzzing, static analysis, dynamic analysis, and model checking.

**Emerging Software Supply Chain (SSC) Attacks.** SSC attacks surfaced recently still leverage some known types of vulnerabilities (i.e., not entirely novel approaches). Their primary focus lies in strategically infiltrating critical SSCs (e.g., submitting vulnerable code changes) to reach a vast number of end-user devices. Reference [21] outlines the 107 unique SSC attack vectors used in 94 real-world attacks or identified vulnerabilities. The notable examples include Heartbleed [29], Log4Shell [29], and SolarWinds [31]. In some cases, developers are directly targeted because their compromised credentials enable attackers to submit malicious code changes. It is exemplified by the CircleCI incident, where a developer laptop was compromised to steal 2FA (two-factor authentication)-backed single sign-on sessions [33], and the LastPass incident, where keys were stolen via key logger malware [34][35].

**Existing SSC Mitigation Techniques.** Reference [21] identified the 33 detective or preventive techniques (e.g., production branch protection [24][25], unused dependency removal [26], version pinning [27], and open source vulnerability scanner integration with CI/CD [23]). Those techniques focus on securing the build process and downstream merging (e.g., known as build reproducibility and bootstrappability[15] [36]).

Recognizing the increasing threat of developer credential theft, GitHub now mandates 2FA for developer accounts associated with critical projects. It implies more widespread industry participation is crucial [32]. Notably, [20] emphasizes the need for community-driven efforts to empower stakeholders in securing the SSC. That includes addressing human factors to reduce developer overwhelm, and systematically decreasing the attack surface for individual software developers. Such efforts could involve establishing usable communication channels across projects and encompassing tools within the build process and CI/CD. However, they offer less emphasis on protecting upstream code check-ins or improving the credibility or security coverage of other vital upstream development activities.

**Faulty Module Prediction.** The concept of software fault prediction shares similarity with the concept of likely-vulnerable code change prediction presented in this study. The field of software fault prediction has been extensively studied in the past [1][5], focusing on identifying modules likely to contain defects. It aligns with the objective of this work that is to predict the potential for vulnerabilities within code changes.

This study differs significantly from those previous works. Most prior research concentrated on predicting general software defects rather than pinpointing vulnerabilities. Furthermore, their prediction granularity typically targeted software modules or subsystems. Thus, those techniques are used in a periodic manner, aiming to select the targets for testing in order to optimize the allocation of limited testing resources. It remains evident in [10], which specifically focuses on vulnerabilities but predicts the most vulnerable modules. In contrast, this paper offers a distinct advantage by providing an online classification for every code change.

Various existing works focused on optimizing features, classifiers, and filters for defect prediction. The past studies have achieved an average detection probability of 71% with a false alarm rate of 25% [1], which was deemed acceptable. Other techniques explored simple heuristics instead of supervised learning, such as [7], which tracks recently modified files, previously buggy files, and their nearby files. Despite continuous efforts to improve prediction techniques [8], attempts to deploy such techniques in commercial software development environments [9] raised skepticisms. It may have been difficult in part because those techniques were neither originally designed for online prediction nor targeting high-priority problems (e.g., high severity security issues).

The long-term accuracy of a defect prediction system (e.g., over multiple years) is significantly influenced by defect triggers. For instance, a high number of bugs found in a file could indicate: either frequent code changes continuously introducing new bugs (assuming consistent triggers), or increased testing efforts uncovering existing bugs over a short period. In the latter case, the buggy modules would be less likely to exhibit new bugs in the near future than the former case. Therefore, to accurately predict the likelihood of a code change introducing new security bugs, it is crucial to capture and analyze both development progress and code change statistics. Such aspects, however, were not the main focus in many previous defect prediction works.

---

[15] Bootstrappable Builds, available at https://bootstrappable.org/

**Vulnerability Prevention Techniques.** Our VP approach shares a common goal with other approaches that focus on enhancing the code review process or providing secure coding education to software engineers. Techniques exist to improve code reviews by identifying syntax errors, common bugs (e.g., through Linting), and coding style issues. Others are to ensure comprehensive unit testing and appropriate reviewer assignments. The underlying motivation of VP aligns with those efforts: to proactively prevent the introduction of vulnerabilities during the development phase.

Vulnerability prevention can be also achieved at the programming language level. Type-safe languages (e.g., Java used in diverse areas including Android framework and app developments) offer inherent protection against common vulnerabilities, such as buffer overflows, integer overflows, and format string vulnerabilities. However, despite efforts to design safer languages like Go, native C/C++ and assembly code remain prevalent, especially in mobile platforms and Internet-of-Things system software (e.g., AOSP). This prevalence leaves those codebases exposed to a wide range of common software vulnerabilities.

**Software Bug Characterization.** Some other previous works have sought to characterize software defects. Those efforts utilize a variety of methods, including code inspection, analyzing previously found bugs, static analysis, and long-term project history analysis. A notable example is [6], a quantitative study based on a static analysis of the Linux kernel commit history data. Additionally, some other works employed software fault injectors that emulate common software defects (e.g., using ODC-based models [15]) to study the resulting consequences [16].

While static analysis can be useful for identifying certain types of known bugs, it faces limitations when it comes to detecting the wide range of realistic security bugs. The high false positive rate and binary decision output (warning vs. no warning) make it challenging to apply static analysis at the granular level of code changes. If the baseline codebase already contains numerous warnings, it becomes difficult to determine whether new warnings resulting from a given code change are actually caused by that code change.

## X. CONCLUSION
This paper presented a practical, preemptive security testing approach that is based on an accurate, online prediction of likely-vulnerable code changes at the pre-submit time. We presented the three types of new feature data that are effective in vulnerability prediction and evaluated their recall and precision via N-fold validation using the data from the large and important Android open source project. We also evaluated the online deployment mode, and identified the subset of feature data types that are not specific to a target project where the training data is collected and thus can be used for other projects (e.g., multi-projects setting). The evaluation results showed that our VP framework identifies ~80% of the evaluated, vulnerability-inducing changes at the pre-submit time with 98% precision and <1.7% false positive ratio. The positive results call for future researches (e.g., using advanced ML and GenAI techniques) to leverage the VP approach or framework for the upstream open source projects managed by communities and are at the same time critical for the numerous software and computer products used by several billions of users in a daily basis.

The urgency of this paper stems from its potential societal benefits. Widespread adoption of ML-based approaches like the VP framework could greatly enhance our abilities to share the credibility data of open source contributors and projects. Such shared data would empower open source communities to combat threats like fake accounts (as seen in the Linux XZ util backdoor attack[16]). Additionally, this ML-based approach can facilitate rapid response across open source projects when long-planned attacks emerge. Sharing information across similar or downstream projects enhances preparedness and reduce response time to similar attacks. Therefore, we call for an open-source community initiative to establish a practice of sharing the credibility database of developers and projects for hardening our open source software supply-chains that numerous computer and software products depend on.

---

[16] FAQ on the xz-utils backdoor (CVE-2024-3094), available at https://gist.github.com/thesamesam/223949d5a074ebc3dce9ee78baad9e27